**Empirical evidence for a double step climate change in twentieth century**

Belolipetsky P.V. [1,2 *], Bartsev S.I.[2], Degermendzhi A.G.[2], Huang-Hsiung Hsu[3], Varotsos C.A.[4]

[1] *Institute of Computational Modelling, SB RAS, Krasnoyarsk, Russia*
[2] *Institute of Biophysics, SB RAS, Krasnoyarsk, Russia*
[3] *Research Center for Environmental Change, Academia Sinica, Taipei, Taiwan*
[4] *University of Athens, Panepistimiopolis, Athens, Greece*

[*]*email: pbel@icm.krasn.ru*

**Abstract**

In this study we used the sea surface temperature (SST), El-Nino southern oscillation (ENSO) and Pacific decadal oscillation (PDO) time-series for the time period 1900-2012 in order to investigate plausible manifestation of sharp increases in temperature.

It was found that the widely observed warming in the past century did not occur smoothly but sharply. This fact is more pronounced at the latitude zone $30^oS$ - $60^oN$ during the years 1925/1926 and 1987/1988. We hypothesise that there were two major climate regime shifts in 1925/1926 and 1987/1988 years. During these shifts the mean value of temperature rises, over which natural variability associated with ENSO, PDO and other factors occurs. During each sharp increase mean SST in tropics/north middle latitudes increased by about 0.28/0.36 °C. Most of other temperature anomalies are explained by ENSO and PDO. The existence of these shifts tends to be masked by natural variability.

This hypothesis has allowed us to develop very simple linear regression models which explain the main features of temperature anomalies from $30^oS$ to $60^oN$ observed in the past century. Additionally, two remarkable outcomes revealed from this analysis. The first one is that linear regression coefficients can be calculated by employing a limited length of the temperature time-series (e.g., from 1910 till 1940) and reproduce quite well the whole time-series from 1900 up to now. The second one is that a good quality of reproduction is achieved by using only two factors (ENSO/PDO and sharp changes for tropics/northern middle altitudes).

## 1. Introduction

The principal indicator of global warming is, by definition, the global mean temperature. The 20$^{th}$ century increase in global mean temperature has been well documented – there was an increase of about 0.75 °C between 1880 and 2008 (Intergovernmental Panel on Climate Change (IPCC) 2007). This warming was not straightforward – the global temperature increased in the first part of the century, then slightly decreased in the years 1940-1970, subsequently increased again and stayed almost flat during the last decade. Additionally, temperatures have always fluctuated rapidly with amplitudes up to 0.5 °C over small time intervals (e.g. years). Both natural and anthropogenic influences have caused the twentieth century climate change, but their relative roles and regional impacts are still under debate (e.g., Kondratyev and Varotsos, 1995; Varotsos et al. 2007).

There are many studies showing pronounced sharp regime shifts in climate and ecological systems occurred in the 20th century (Trenberth and Hurrel 1994; Deser et. al 2004; Yasunaka and Hanawa 2002). In this regard, Yasunaka and Hanawa (2002) described a "regime shift" as an abrupt transition from one quasi-steady climatic state to another, and its transition period is much shorter than the lengths of the individual epochs of each climatic state. Different studies mark out existence of many shifts in 20th century. Yasunaka and Hanawa (2002) detected six regime shifts in northern hemisphere SST field in the period from 1910s to the 1990s: 1925/1926, 1945/1946, 1957/1958, 1970/1971, 1976/1977 and 1988/1989. Deser et. al (2004) considered shifts in 1925, 1946 and 1976 years.

In our study only shifts in 1925/1926 and 1987/1988 years are considered. One of the reasons is that our working definition of shifts has some differences from that used by Yasunaka and Hanawa, Deser et. al and many others. We herewith define climate regime as a quasi-steady state with known sources of variability. Additionally, a climate regime shift is significant and systematic changes, which occur besides intra regime variability and separating one climate regime from another. In other words, the two cases of the mean value of temperature increase (in the years

1925/1926 and 1987/1988), take place over the existed natural variability that is mainly associated with ENSO and PDO. It should be noted that sometimes the abrupt increases in the mean value of temperature is almost masked by natural variability. For example, step change of SST in the tropics in 1976 is clearly seen in time series but the shift in 1987 is not obvious at all (Fig. 1). But the above-said shift in 1976 is, in general, associated with ENSO and could be almost reproduced by direct linear association with ENSO Nino34 index (Fig. 1b). Therefore, according to our definition, it should not be considered as a regime shift, because it is described by known intra-regime variability.

## 2. Data and methodology

All the calculations presented below were made in Excel by means of standard functions. A reconstruction of monthly mean surface temperature anomalies, $T_R$, from input parameters is performed by employing the following relationship:

$$T(t) = c_o + \sum_{i=1}^{n} c_i \cdot X_i(t - \Delta t_i).$$

Here $X_i$ stands for the climate factors; $\Delta t_i$ are the time lags in months; $c_i$ - the fitted coefficients and n – the number of variables in regression analysis. The fitted coefficients are obtained by standard Excel function for multiply linear regression.

All of the datasets used are freely available at Climate Explorer site (climexp.knmi.nl). In the analysis presented we used several reconstructions of the observed sea surface temperature anomalies - HadSST2 and Reynolds v2. Our choice of 1900 as the starting year for the analysis is based upon data coverage considerations. For the period from 1900 up to the beginning of the 1980s we used HadSST2 and later Reynolds v2 as SST anomalies (as we thought that remote sensing data have more precise spatial coverage). Climate regimes, ENSO and PDO were considered as factors determining observed temperature anomalies. As a proxy for ENSO we assumed the Nino34 index (obtained from HadISST1), and used it with one month lag. For the PDO we used reconstruction from ERSST with no time lag. All datasets used were prepared and downloaded from Climate Explorer site (climexp.knmi.nl).

## 3. Results

Consider tropical SSTs (30°S-30°N) (Fig. 1 blue line). As it is shown in this figure most of the SST variability may be explained by ENSO. At first sight these natural oscillations should be accompanied by some continuous warming. It is worth to be noted, however, that most of variability from middle 80s to the present could be reproduced quite well by linear regression on ENSO without a warming trend (Fig. 1a red line). On the other hand regression on ENSO reproduces the period from 1950 till middle 80s (Fig. 1b red line) quite well, but is inadequate later.

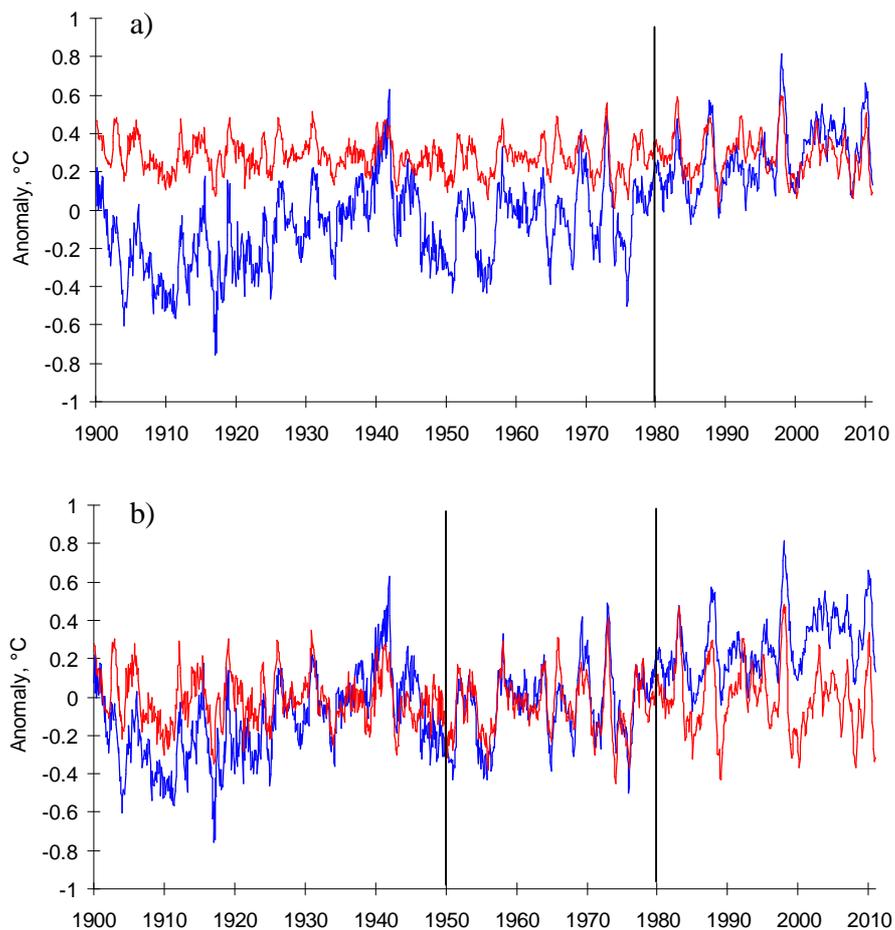

Fig. 1. Blue line - observed SST in tropics (30°S-30°N), red line - linear regression on ENSO studied a) by 1984-2010 years; b) by 1950-1980 years.

This suggests that there might be a some sharp change of mean temperature over which ENSO oscillations occurs. For instance, it may be associated with some climate regime shift somewhere in the middle 80th. Thus let us assume that there is another determining climate factor - climate regime index, a step function witch equals zero before shift and equals one after. Validation procedure shows that the best time for this hypothesized regime shift is 1987. In this case temperature anomalies are reproduced without continuous warming at least from 1940s (Fig. 2). If we assume that shift happened at some other time then reconstructed temperature anomaly will appear visually obvious discontinuity which is not observed in the data.

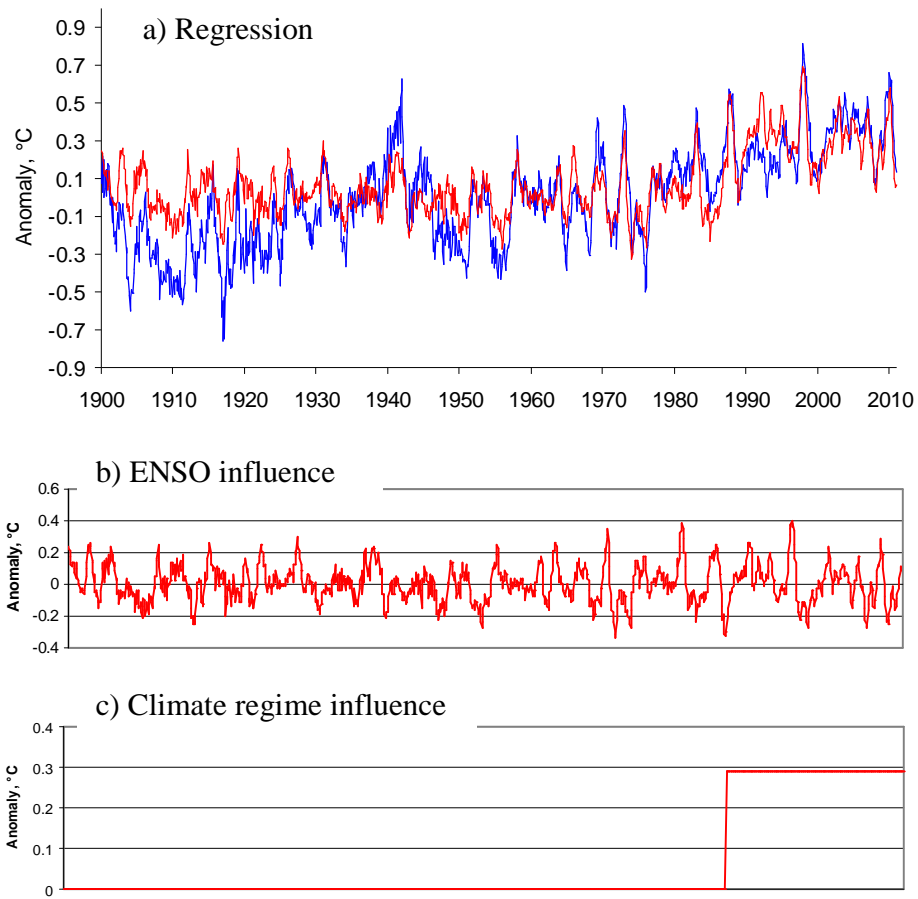

Fig. 2. a) Blue line - SST in tropics, red line - linear regression on ENSO and climate regime, studied by 1950-2012 years; b) ENSO influence on tropical SST; c) climate regime influence on tropical SST.

In order to get an adequate reconstruction for the whole period let's assume that there was another shift of the same magnitude in the first part of the century. Validation shows that it could occur between 1925 and 1926. In such a way the climate regime index became two step function, which equals -1 before 1926, 0 between 1926 and 1987 and 1 after. It is worth noting here that the values -1, 0, 1 are indicators of different regimes with the assumption that shifts in 1925/1926 and 1987/1988 produce equal changes in temperature anomalies. Using this model adequate reconstruction from 1900 to the present is obtained (Fig. 3). Of course, there are some differences between observation and reconstruction. But we do not consider many other possible influencing factors such as volcanoes, solar activity etc. Our main aim is to mark out the main factors allowing for adequate reconstructions.

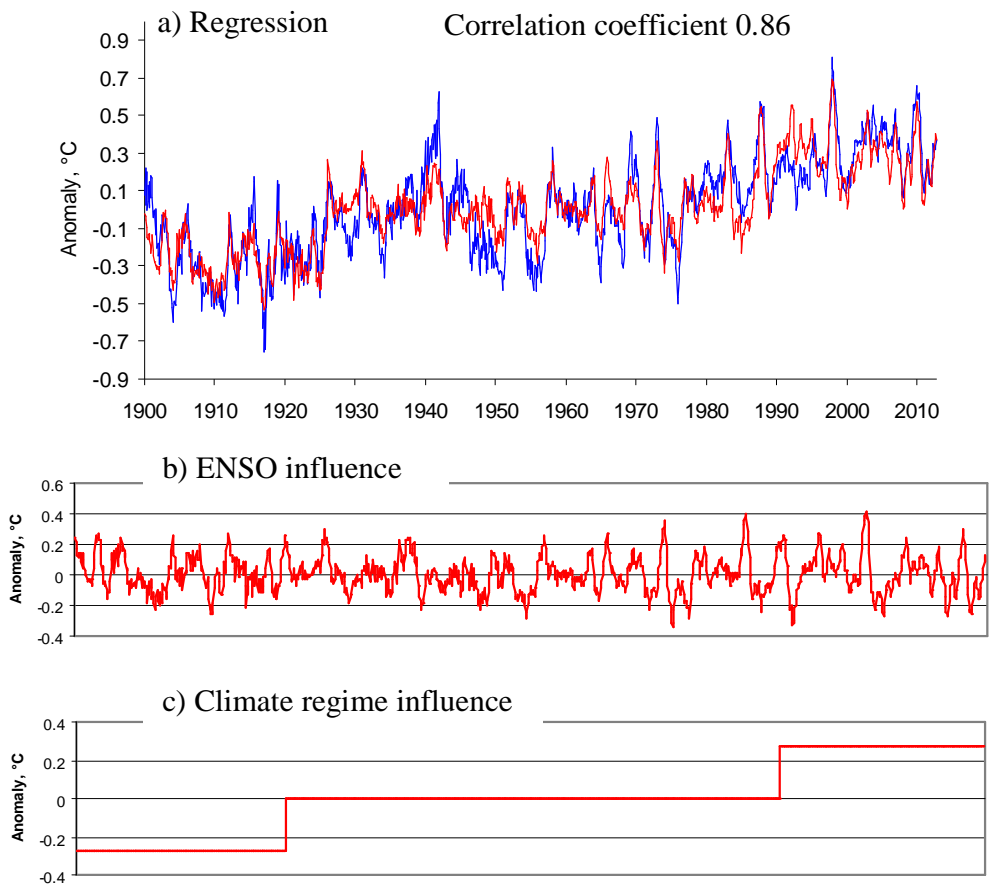

Fig. 3. a) Blue line - SST in tropics, red line - linear regression on ENSO and climate regime, studied by 1900-2012 years b) ENSO influence on tropical SST; c) climate regime influence on tropical SST.

Let us now perform the same linear regression analysis for middle latitude SST, as we did for the tropics. Here instead of ENSO most of the variability is closely associated with PDO. For a good reconstruction in this region climate shifts would have happened later than in the tropics - in the middle of 1926 instead of the 1925/1926 boundary, and in the first part of 1988 instead of middle 1987. Again SST anomalies are reproduced quite well (Fig. 4). These dates have been obtained by the same calibration procedure as for the tropics. Dates have been obtained by the same validation procedure as in tropics - other times where shifts occur, lead to visually obvious discontinuity in reconstructed temperature anomaly, which is not observed in the real data.

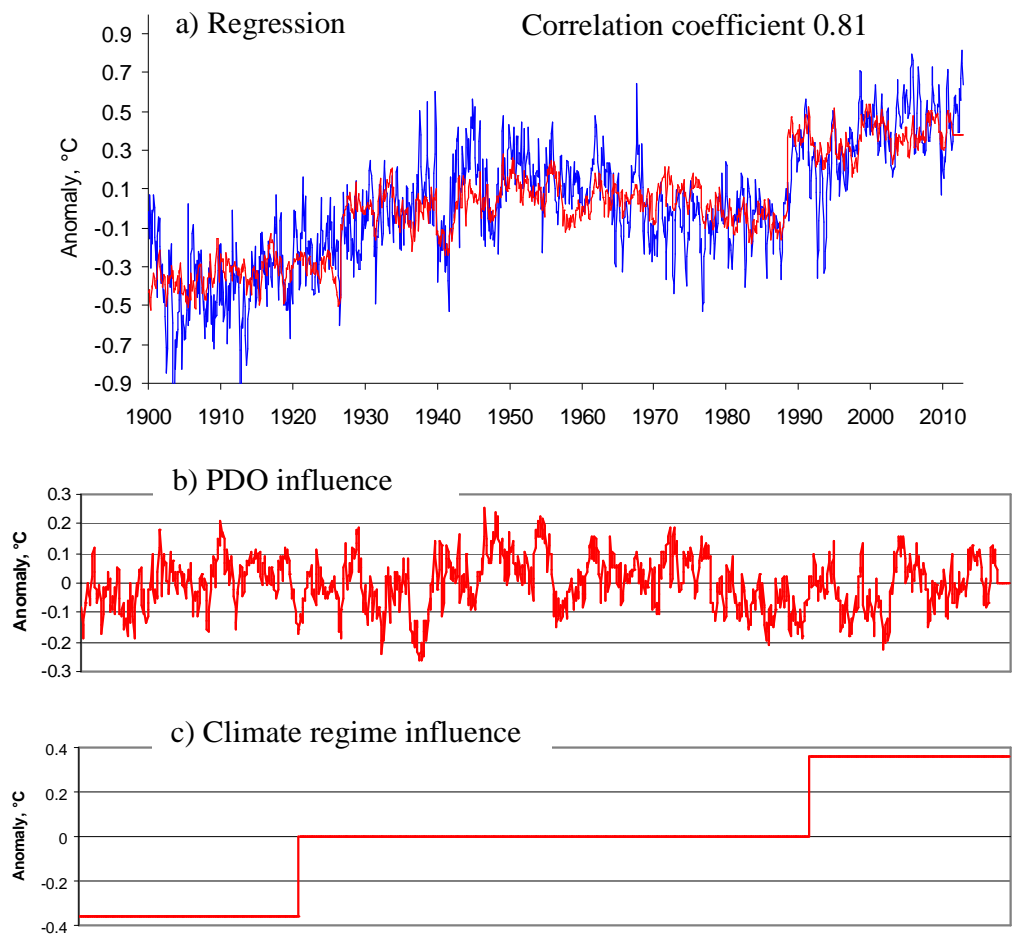

Fig. 4. a) Blue line - SST in north middle latitudes (30$^o$N-60$^o$N), red line - linear regression on PDO and climate regime, studied by 1900-2012 years b) PDO influence on SST in this region; c) climate regime influence on SST in this region.

A quite remarkable feature is that linear regression coefficients can be fitted simply by using the data from 1910 till 1940 (15 years to both side from shift in 1925/1926) and with almost the same quality reproduce the whole period from 1900 till now (Fig. 5). This shows the robustness of our suggested relationships. Another piece of evidence for robustness is that the consideration of these shifts separately (e.g. using two climate regime indexes in linear regression - one with shift in 1925/1926 and another in 1987/1988) gives nearly the same amplitudes for the shifts - 0.26 °C and 0.28 °C for the tropics, 0.38 °C and 0.34 °C for northern middle latitudes. The assumption that shifts in 1925/1926 and 1987/1988 produce equal changes to SST regression, shows 0.28 °C temperature rise in tropics and 0.36 °C rise in northern middle latitudes.

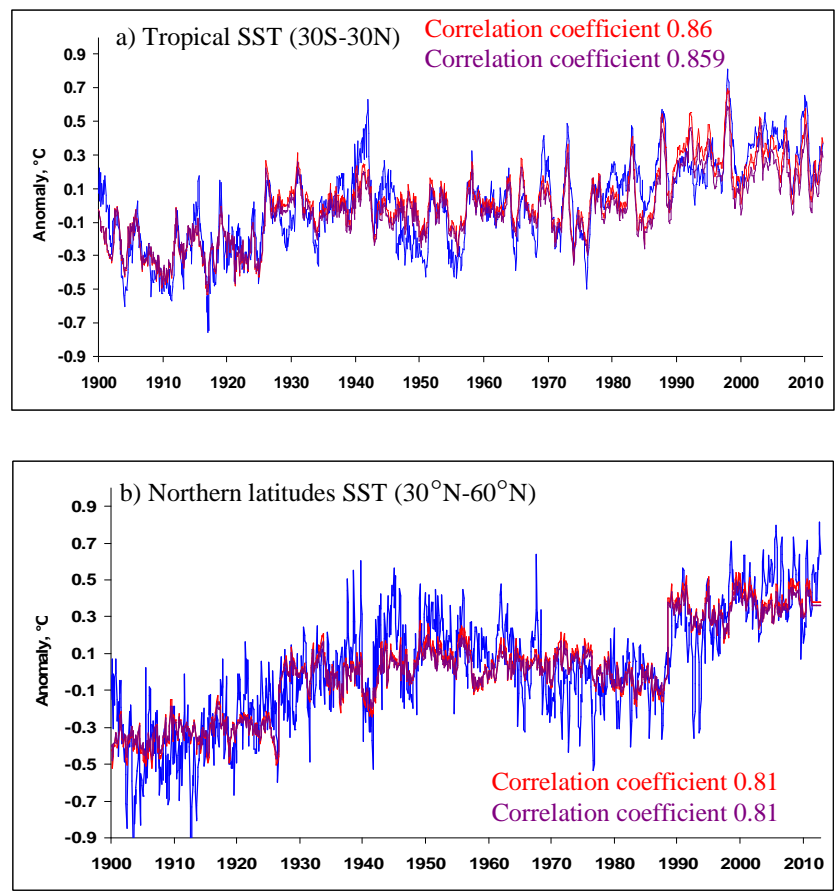

Fig. 5. a) Blue line - SST in tropics (30°S-30°N), red line - linear regression on ENSO and climate regime with training period 1900-2012 years, purple line - the same linear regression with training period 1910-1940 years; b) the same as "a" but for north middle latitudes (30°N-60°N).

## 4. Discussion

Lean and Rind (2008) performed multivariate linear regression analysis of the natural and anthropogenic influences on global surface temperature anomalies. They concluded that much of the variability in global climate arises from processes that can be identified and their impact on the global surface temperature quantified by direct linear association with the observations. The main part of natural variability in analysis performed by Lean and Rind (2008) is associated with ENSO. However, the influence of ENSO is clearly seen only in the tropics (30°S-30°N) while in the middle latitudes and polar regions this influence is not so significant.

One of the ways to deal with this is to consider different parts of Earth surface separately. We used this approach and the result obtained is the hypothesis that at least in two regions (tropics (30°S-30°N) and northern middle latitudes (30°N-60°N)), the conditional warming was not continuous process, but occurred through sharp regime changes in the years 1925/1926 and 1987/1988. For each region we have developed very simple linear regression models representing temperature dynamics without a continuous warming process. This revealed that the dynamics of the monthly average tropical SST anomaly could be adequately reproduced by two factors - ENSO variability (Nino 3.4 index) and two climate shifts in 1925/1926 and 1987/1988 years. Northern middle latitudes SST temperature anomaly could be reproduced in general by the same factors, except that ENSO is changed to PDO here. Continents in these parts appear the same dynamics, but with enhanced variability. The results enable us to suggest a hypothesis that the warming observed from 30°S to 60°N was actually not continuous, but a step function.

What do observational studies suggest about the shifts in 1925/1926 and 1987/1988? There are many pieces of evidence for climate and ecological regime shifts in the considered years. Shifts were observed in combined physical and biological variables (Hare and Mantua, 2000; de Young et. al 2004), birds populations (Veit et. al 1996), fish populations (Chavez et. al 2003), local ecosystems (Tian et. al 2008, Beaugrand and Reid 2003), global carbon cycle (Sarmiento et. al 2010), pressure patterns (Trenberth and Hurrel 1994; Deser et. al 2004) and temperature anomalies patterns (Yasunaka and Hanawa 2002; Lo and Hsu 2010). Reviewing these studies we seek for evidence of the existence of shifts in 1925/1926 and 1987/1988 years and of their differences from other observed shifts.

It is now widely accepted that a climatic regime shift revealed in the North Pacific Ocean in the winter of 1976-1977. Hare and Mantua (2000) assembled 100 environmental time series, 31 climatic and 69 biological, to determine if there is evidence for common regime signals in the 1965-1997 period of record. Their analysis reproduces previously documented features of the 1977 regime shift, and identifies a further shift in 1988/1989 in some components of the North Pacific ecosystem. A notable feature of the 1988/1989 regime shift is the relative clarity that it has been observed in biological records, which is in contrast to the relative lack of clear changes denominated by indices of Pacific climate. This may be assumed as indication that some substantial climate changes occurred besides known sources of variability. For example, the discovery of so-called biological regime shifts in 1976-1977 preceded the description of the underlying physical variability. A decade or more after the observations of sardine and anchovy variations, scientists discovered fluctuations in air and ocean temperatures, and in atmospheric circulation that were remarkably similar in phase and duration to the biological records (Chavez et. al 2003). It takes even more time to show that sardine and anchovy variations governed not by their interactions but fully by physical variability (Sugihara et. al 2012). In this study any underlying physical processes for 1988 shift are represented by means of the climate regime index.

Yasunaka and Hanawa (2002) applied an empirical orthogonal function (EOF) analysis to northern hemisphere SST field and detected, as mentioned above, six regime shifts in the period from 1910s to the 1990s. They highlighted the difference of 1925/1926 and 1987/1988 regime changes from other shifts - "According to spatial pattern correlation between SST difference maps of regime shifts, it is found that the 1945/1946, 1957/1958, 1970/1971 and 1976/1977 regime shifts are similar pattern, while the 1925/1926 and 1988/1989 regime shifts are somewhat different." Our hypothesis (which we believe we have shown to be viable) is that the 1945/1946, 1957/1958, 1970/1971 and 1976/1977 regime shifts were closely associated with changes in ENSO and PDO and are therefore described by known intra regime variability.

Lo and Hsu (2010) investigated extra-tropical Northern Hemisphere temperature anomalies and suggested that the main reason for recently observed warming in these latitudes is a climate shift in 1987, which is in accordance to our hypothesis in this study. In particular, they found unprecedented from early 1940s phenomenon in the late 1980s - temperature fluctuation synchronization in widespread areas of Northern Hemisphere (Fig. 6).

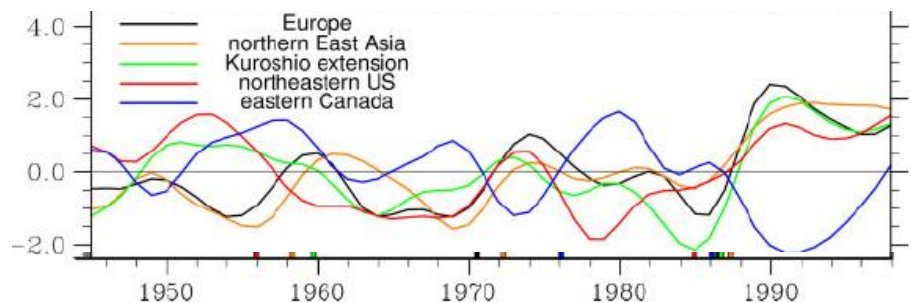

Fig. 6. Time series of 9-year running-mean surface temperature anomalies (°C) in five chosen regions. Modified from Lo and Hsu (2010).

Therefore, the linear regression analysis discussed in this study showed the occurrence of the climate shifts in 1925/1926 and 1987/1988. The way in which the existence of these shifts was noticed, was recently described in more detail (Belolipetsky and Bartsev 2012). It should be mentioned that detecting the possibility of their existence is not obvious. In the first place one should consider SST in order to detect the shifts because highly variable land temperatures mask the effect. Second, different latitude bands should be considered separately, because mixing different sources of variability also smoothes the effect. Third, consideration of yearly averaged values instead of monthly averages will make the detection of shifts more difficult.

However, we wish to make it clear that detection of a regime shift is much easier than understanding the process or processes determining it. So we are not speculating here about physical mechanisms and reasons for shifts. There are many possible variants as climate is complex nonlinear dynamical system. The reasons may be intrinsic causes, some indirect solar or volcanic forcing, or result of anthropogenic forcing. Our aim is to show the possibility and evidence for the hypothesis that observed warming in latitudes from 30°S to 60°N occurred not continuously but by means of sharp changes.

## 5. Conclusions

The linear regression models presented above may capture most of the critical dynamics of the surface temperature records from 30°S to 60°N. It should be emphasised that if we accept two climate shifts (those occurred in 1925/1926 and in 1987/1988) the principal features of the observed SST temperature anomalies from 30°S to 60°N in the past century could be very easily explained. In addition, there are many independent indications that depict that these two shifts are real phenomena (Veit et. al 1996; Chavez et. al 2003; Hare and Mantua, 2000; de Young et. al 2004; Tian et. al 2008; Sarmiento et. al 2010; Yasunaka and Hanawa 2002, Lo and Hsu 2010 and others).

From the discussion above follows that there are two remarkable outcomes. The first finding is that the linear regression coefficients deduced from a small part of data (e.g. from 1910 till 1940) can reliably reproduce the entire data set (i.e., from 1900 to the present). The second one is that good quality of the reproduction may be achieved by using only two factors (i.e., ENSO/PDO and shifts for tropics/northern middle latitudes).

**Acknowledgments.** We thank V.M. Belolipetsky and Robin Edwards for fruitful discussions and suggestions, which helped substantially to improve the results and manuscript.

## References

Beaugrand, G., & Reid, P. C. (2003). Long-term changes in phytoplankton, zooplankton and salmon linked to climate. Global Change Biology, 9, 801–817.

Belolipetsky PV, Bartsev SI (2012) Hypothesis About Mechanics of Global Warming from 1900 Till Now. Preprint. viXra:1212.0172.

Chavez FP, Ryan J, Lluch-Cota SE, Miguel Niquen C (2003) From Anchovies to Sardines and back: multidecadal change in the Pacific Ocean. Science, 299, 217-221.

Deser C, Phillips AS, Hurrell JW (2004) Pacific Interdecadal Climate Variability: Linkages between the Tropics and the North Pacific during Boreal Winter since 1900. Journal of Climate, 17, 3109–3124.

deYoung B, Harris R, Alheit J, Beaugrand G, Mantua N, Shannon L (2004) Detection regime shifts in the ocean: data considerations. Progress in Oceanography, 60, 143-164.

Hare SR, Mantua NJ (2000) Empirical evidence for North Pacific regime shifts in 1977 and 1989. Progress in Oceanography, 47, 103-145.

Intergovernmental Panel on Climate Change (2007) Climate Change 2007: The Physical Science Basis, Contribution of Working Group I to the Fourth Assessment Report of the Intergovernmental Panel on Climate Change, edited by S. Solomon et al., Cambridge Univ. Press, Cambridge, U. K.

Kondratyev, K. Y. and Varotsos C (1995) Atmospheric greenhouse effect in the context of global climate-change, Nuovo Cimento della Societa Italiana di Fisica C-Geophysics and Space Physics, 18(2), 123–151.

Lean JL, Rind DH (2008) How natural and anthropogenic influences alter global and regional surface temperatures: 1889 to 2006. Geophys. Res. Lett., 35, L18701, doi:10.1029/2008GL034864.

Lo TT, Hsu HH (2010) Change in the dominant decadal patterns and the late 1980s abrupt warming in the extratropical northern hemisphere. Atmospheric Science Letters, 11, 210–215.

Sarmiento JL, Gloor M, Gruber N, Beaulieu C, Jacobson AR, Mikaloff Fletcher SE, Pacala S, Rodgers K (2010) Trends and regional distributions of land and ocean carbon sinks. Biogeoscinces, 7, 2351-2367.

Sugihara G, May R, Ye H, Hsieh C, Deyle E, Fogarty M, Munch S. (2012) Detecting Causality in Complex Ecosystems, Science, 338, 496-500.

Trenberth KE, Hurrell JW (1994) Decadal atmosphere-ocean variations in the Pacific. Climate Dynamics, 9, 303.

Tian Y, Kidokoro H, Watanabe T, Iguchi N (2008) The late 1980s regime shift in the ecosystem of Tsushima warm current in the Japan/East Sea: Evidence from historical data and possible mechanisms. Progress in oceanography, 77, 127-145.

Varotsos C, Assimakopoulos MN, Efstathiou M. (2007) Technical note: long-term memory effect in the atmospheric $CO_2$ concentration at Mauna Loa. Atmospheric Chemistry and Physics. 7(3), 629–634.

Veit RR, Pyle P, McGowan JA (1996) Ocean warming and long-term change in pelagic bird abundance within the California current system. Marine ecology progress series, Vol. 139, 11-18.

Yasunaka S, Hanawa K (2002) Regime shifts found in Northern Hemisphere SST Field. Journal of meteorological society of Japan, Vol. 80, No. 1, pp. 119-135.